\documentclass
[aps,twocolumn,amsmath,amssymb,showpacs,floatfix,superscriptaddress]
{revtex4}
\usepackage{color,graphicx}
\usepackage{mathptmx}                % math+times fonts
\usepackage{dcolumn}                 % Align table columns on decimal point
\usepackage{bm}                      % bold math
\begin{document}

\title{Pairing effects on neutrino transport in low-density stellar matter}  

\author{S.Burrello}
\affiliation{INFN-LNS, Laboratori Nazionali del Sud, 95123 Catania, Italy}
\affiliation{Dipartimento di Fisica e Astronomia, Universit\`a di Catania, 95123 Catania, Italy}
%\affiliation{INFN-LNS, Laboratori Nazionali del Sud, 95123 Catania, Italy}
%\affiliation{INFN, Laboratori Nazionali del Sud and Dip. di Fisica e Astronomia, Universit\`a di Catania, 95123 Catania, Italy}
%%%  + Dipartimento di Fisica%%%  + Dipartimento di Fisica
\author{M.Colonna}
\affiliation{INFN-LNS, Laboratori Nazionali del Sud, 95123 Catania, Italy}
\author{F.Matera}
\affiliation{Dipartimento di Fisica e Astronomia, and Istituto Nazionale di Fisica Nucleare, 50019 Sesto Fiorentino, Firenze, Italy}

\begin{abstract}
We investigate the impact of pairing correlations on neutrino transport in 
stellar matter. Our analysis is extended to nuclear matter conditions
%density and charge asymmetry conditions
%{\bf of nuclear matter,} 
where large density fluctuations develop, associated with the onset of the liquid--vapor
phase transition, and clustering phenomena occur. 
Within a thermodynamical treatment, we show that at moderate temperatures, 
where pairing effects are still active, the %!! neutral current neutrino scattering 
scattering of neutrinos in the nuclear medium
is significantly affected by pairing correlations, which increase the neutrino trapping,
thus modifying the cooling mechanism, by neutrino emission, of neutron stars.
%Important consequences on the cooling mechanism, by neutrino emission, of neutron stars can be envisaged. 
\end{abstract}

\date{\today}

%\pacs{XXXX}
%\pacs{05.70.Fh, 21.30.Fe, 21.65.Ef, 74.20.Fg}

\maketitle

%###################################

%%%%%%%%%%%%%%%%%%%%%%%%%%%%%%%%%%%%%%%%%%%%%%%%%%%%%%%%%%%%%%%%%%%%%%
Transport properties of neutrinos play an essential role in
the description of  gravitational  collapse,  supernovae,  protoneutron
stars and binary mergers of compact objects.
For instance, the cooling process of newly formed neutron stars %young
is, over a broad time interval 
($10^{-10}$ - $10^5$ yr), dominated by the emission of neutrinos of all flavors.
%emission, due to several microscopic processes. 
%Thus, neutrinos of  all  flavors  emitted  from  newly  formed  neutron  stars  in
%supernova  explosions are  
In fact, the neutrino flux is the  only  direct  probe  of  the
mechanism of supernovae and the structure of protoneutron
stars \cite{Be85,Bu90,Pra01,Ja07,Bal14}.  

The  most  important  ingredient  of  neutrino  transport
calculations  is  the  neutrino  opacity,
%linked to the neutrino mean-free path in the stellar matter.
%Both 
essentially determined by charged current absorption and neutral current scattering  
reactions.  
%are  important  sources  of  opacity. 
%[Neutral
%current  processes  involve  all  flavors  of  neutrinos  scattering
%on  baryons  and  leptons.]  
%While  scattering  from  electrons  is
%important for energy and momentum transfer in the process
%of thermalizing neutrinos, for both energy and lepton number
%transport,  neutrino-baryon  scattering  and  absorption  are  the
%dominant processes \cite{Re98,Bu98,Re00}.
While  scattering  from  electrons  
dominates the neutrino thermalization process, 
neutrino-baryon  scattering  and  absorption are the 
leading mechanisms for  energy and lepton number
transport \cite{Re98,Re99,Bu98,Re00}.

Recently, %an enormous effort has been done 
many efforts have been devoted
to describe the neutrino production and
interactions in great detail \cite{Wa03,Ho04,Ma12,Pa14}.
In particular, it appears that this mechanism is clearly affected by
general phenomena, such as phase transitions, characterizing the
behavior of interacting many-body systems. 
Liquid-gas phase transitions are expected to appear
for nuclear matter at sub-saturation density and relatively low temperature ($T \lesssim 15$ MeV), 
driven by the unstable nuclear mean-field (i.e. by spinodal instabilities) \cite{rep04}.   
This important feature of nuclear matter is closely connected to the  
multifragmentation mechanism
experimentally observed in nuclear reactions \cite{Borderie}
and to the occurrence of clustering phenomena in the inner crust of neutron stars
\cite{Lattimer,Camille}.    
It was recently pointed  out  that  large  density  fluctuations,
associated  with  the  first  order  nuclear  liquid-gas  phase  transition, hugely increase the scattering of neutrinos
\cite{Marg2004}, thus quenching their emissivity processes in low
density regions. %, which are encountered in the crust of a neutron star.

As a quite general feature, fermionic systems may also exhibit 
pairing correlations. Their importance is widely discussed in the
astrophysical context, as far as cooling processes \cite{Pa09} 
and glitch phenomena \cite{Ca00,Se05} are concerned.   
%\textcolor{red}{For instance, the cooling or thermalization
%time of the crust of neutron stars (defined as the time needed for the crust matter
%to equilibrate its temperature to that of the colder
%core) has a crucial dependence of the heat capacity of the stellar matter, 
%which in turn is strongly influenced by pairing correlations.}

%\textcolor{red}{It should be noticed  that} \textcolor{blue}
Since pairing correlations are mostly active at 
low density and relatively low temperature, below the critical temperature
for the transition from superfluid to normal matter, %\textcolor{red}{. Thus is} \textcolor{blue}
in a certain 
region of the nuclear matter phase diagram volume instabilities may co-exist
with strong pairing effects \cite{Bur14}. 
%[Indeed, 
%in recent studies, it was observed that pairing correlations 
%can have a significant impact
%on some relevant features associated with spinodal instabilities in asymmetric matter.]

The aim of this work is to investigate the influence of pairing correlations
on the neutrino scattering cross section, in conditions of moderate temperature, charge asymmetry and 
low density (close to the spinodal border), which are 
encountered in neutron stars,
as well as in supernova explosions (pre-bounce phase). 
%which are close to the spinodal border. 
%where spinodal instabilities occur. 

%inside a thin layer of the protoneutron star
%(PNS) for  densities  in  the  range  between  about  0.1 r0
%and  0.6 r0, where r0
%is the nuclear matter density at saturation. The re-
%sulting  increase  in  the  pressure  may  induce  strong  particle
%ejection a few hundred milliseconds after the bounce of the
%collapse 1, contributing to the revival of the shock wave.

%\textcolor{blue}{This spinodal border, defined as the plane on which the speed of sound vanishes%, both in symmetric and asymmetric matter, delimits a region of instability characterized by}
%%The onset of the liquid-gas co-
%%existence  phase  is  associated  with  the  
%\textcolor{red}{Spinodal instabilities are related to} a negative curvature in the nucleonic fre%e energy
%density\textcolor{red}{, both in symmetric and asymmetric matter, so that the speed of sound van%ishes at the crossing with the spinodal border}.
At the onset of spinodal instabilities, the speed of sound vanishes and the  nucleonic free energy
density exhibits a negative curvature,  both in symmetric and asymmetric matter.
%\textcolor{green}{[It is not clear for me!]} 
%The analysis
%of the behavior of the thermodynamic potential show that the
%phase transition is first order both in symmetric and asymmetric
%nuclear matter. 
%[Though a unique spinodal instability is expected, 
%the asymmetry induces a change in the isotopic composition
%of the density oscillations, the so-called distillation mechanism,
%and so on the isotopic properties of the clusters which finally emerge. 
%The latter effect is linked to the isovector terms of the nuclear effective
%interaction and to the symmetry energy of the nuclear matter 
%equation of state, on which many investigations are concentrated
%nowadays.]
%do not modify its properties up to very large
%as
%However, 
%The spinodal instability is globally isoscalar-like %\textcolor{green}{[Maybe we should to explain it better!]}
% be-cause the interaction between protons and neutrons is attrac-
%tive at low density. 
%Hence, in the density region close to (or inside) the
%spinodal, 
Owing to the global isoscalar-like character of the instability,
the  density  response  function  is  enhanced,
thus modifying the neutral current neutrino opacity, whereas 
charge current  absorption processes are not affected. 
%\textcolor{red}{This should lead to  a reduced total neutrino mean-free path.}
Hence in the following we will concentrate only on neutral current neutrino
scattering. 
%%%%%%%%%%%%%%%%%%%%%%%%%%%%%%%%%%%%%%%%%%%%%%%%%%%%%%%%%%%%%%%%%%%%%%%%%%%%%%%%%%%%%%%%%%%%%%%%%%%%%%%%%%%%%%%
%%%%%    fin qui
We will consider nonrelativistic nucleons coupled to neutrinos through only
the vector neutral current, neglecting contributions from the axial current, as 
spin--density fluctuations do not present any singular behavior for the assumed 
physical conditions. 
%Thus the weak interaction Lagrangian density reads \cite{Bu98,Iw82}: 

%\begin{equation}
%\mathcal{L}_I (x) = \frac{G_F}{\sqrt{2}}  \bar{\psi}^{(\nu)} (x) \gamma_{0} 
%( 1 - \gamma_5) \psi^{(\nu)}(x) 
% J^{(N)0} (x),
%\end{equation}
%%\begin{equation}
%%j_{\mu}^{(\nu)} (x) = \bar{\psi}^{(\nu)} (x) \gamma_{\mu} ( 1 - \gamma_5) \psi%^%{(\nu)}(x)  
%%\end{equation}
%where $G_F$ is the weak coupling constant, $ \psi^{(\nu)}(x) $ denotes the neut%rino field and 
%\begin{equation}
%J_{0}^{(N)} (x) = \sum_{i = n, p} c_V^{(i)} \rho_i (x),
%\end{equation}
%with $c_V^{(n)} = -0.5$, $c_V^{(p)}$ = 0.036 and 
%$ \rho_i( x)$  (i = n, p) is the nucleon (neutron or proton) local density
% (we use units such that $\hbar=c=k_B=1$). 
%\textcolor{green}{[Should we also clarify what are the $\gamma^0$, $\gamma^5$ matrices?]}  

Then the differential cross section (per unit of volume $V$) for scattering in the medium of neutrinos
with energy $E_{\nu}$,
as a function of the neutrino final energy $E'_{\nu}$ and scattering angle $\theta$,
is given by \cite{Bu98,Iw82}:

\begin{equation}
\frac{1}{V}\frac{d^3 \sigma}{dE'_{\nu} d\Omega^2} = 
\frac{G_F^2}{8 \pi^3}(E'_{\nu})^2(1 + \cos \theta) S_V^{00} (\omega, \mathbf{q}),
%\frac{G_F^2}{32 \pi^2} \frac{E'}{E} L_{\alpha \beta} (k, k') S^{\alpha \beta} (%\omega, \mathbf{q})
\end{equation}
where $\omega = E_{\nu} - E'_{\nu}$ denotes the energy transfer to the medium, 
$\mathbf{q}$ is the momentum transfer, related to $\omega$ and to the neutrino
scattering angle $\theta$  (we use units such that $\hbar=c=k_B=1$). $G_F$ denotes the weak coupling constant and
%\begin{equation}
%k^{\mu} = (\mathbf{k}, E), \qquad k'^{\mu} = (\mathbf{k'}, E')
%\end{equation}
%\begin{equation}
%E = |\mathbf{k}| \qquad E' = |\mathbf{k'}|
%\end{equation}
%\begin{equation}
%\mathbf{q} = \mathbf{k} - \mathbf{k'} \qquad \omega = |\mathbf{k}| - |\mathbf{k'}|
%\end{equation}
$S_V^{00}$ identifies the dynamic form factor, which can be expressed in terms
of the nucleon density-density correlation factor as:
\begin{equation} 
S_V^{00} (\omega, \mathbf{q}) = \int dt\,d\mathbf{r}\, e^{i\omega t}\, e^{-i\mathbf{q}\cdot\mathbf{r}} \langle J^{(N)0} (t, \mathbf{r})J^{(N)0} (0, 0) \rangle,
\end{equation}
where $J_{0}^{(N)} (t, \mathbf{r}) = \sum_{i = n, p} c_V^{(i)} \rho_i (t, \mathbf{r}),$
%\end{equation}
with $c_V^{(n)} = -0.5$, $c_V^{(p)}$ = 0.036.
$ \rho_i(t, \mathbf{r})$  (i = n, p) is the nucleon (neutron or proton) local density.
% (we use units such that $\hbar=c=k_B=1$). 
%\textcolor{green}{[Should we also clarify what are the $\gamma^0$, $\gamma^5$ matrices?]}$.  
Based on energy conservation, we see that  a
typical energy transfer $\omega$
from the medium to the neutrino is
of the order of
$q$
times the thermal velocity  of nucleons $(T/m)^{1/2}$.
Thus,
in the limit of heavy nucleons, when we integrate the differential cross section
over a range of
$\omega$ values,
the other factors in the
integrand can be evaluated at
$\omega = 0$, so that:   %We  can  express this as:
 \begin{equation}
S_V^{00} (\omega,\mathbf{q}) \rightarrow  \delta(\omega) \int d\omega' S_V^{00} (\omega',\mathbf{q}) = 
2 \pi \delta(\omega)S_V^{00}(\mathbf{q}),
\end{equation} 
where $S_V^{00}(\mathbf{q})$ is the static structure factor, which, according to
Eq.(2), corresponds to $n-n$, $p-p$ and $n-p$ density fluctuation correlations taken at equal time:
\begin{equation}
S_V^{00}(\mathbf{q}) = \langle \delta J^{(N)\,0} (\mathbf{q})\delta J^{(N)\,0} (-\mathbf{q}) \rangle .
%\sim \sigma_i^2 = \langle \delta \rho_i \delta \rho_i \rangle
\end{equation} 
Exploiting the  fluctuation-dissipation theorem \cite{Landau} and
%\bibitem{Landau}L.D. Landau, E.M. Lifshitz, Statistical Physics Part 1.
%Vol. 5 (3rd ed.), Butterworth-Heinemann, ISBN 978-0-
%750-63372-7 (1980)
neglecting quantum fluctuations, the static structure factor can be expressed, for a system at temperature T, as:
\begin{equation}
S_V^{00}(\mathbf{q}) = T~\left[ {c_V^{(n)}}^2 \mathcal{C}_{nn}^{-1} (q) + {c_V^{(p)}}^2 \mathcal{C}_{pp}^{-1} (q) + 2 c_V^{(n)}c_V^{(p)}\mathcal{C}_{np}^{-1} (q)\right], 
\end{equation}
where the matrix  $\mathcal{C}^{-1}$ is the inverse of curvature matrix of 
the system free energy density \cite{Bu98}. 

In this paper, we consider
stellar matter where the proton charge is globally 
neutralized by a Fermi gas of electrons.
The local energy density, which is a function of the total density $\rho$
$=\rho_n + \rho_p$, the 
proton fraction $y_p = \rho_p / \rho$ and the
electron density $\rho_e$, can be written as: 
\begin{equation}
\mathcal{E}_{tot}(\rho,y_p,\rho_e) = \mathcal{E}_{NM} +  \mathcal{E}_{NM}^{f} + 
\mathcal{E}_{Coul} + \mathcal{E}_e(\rho_e) ,
\end{equation}
where $\mathcal{E}_e$ is the energy density associated with the electron kinetic energy
%In the equation above,
and the contributions of the Coulomb term, $\mathcal{E}_{Coul}$, related to the interaction
between all charges (protons and electrons), and of nuclear matter surface
terms, $\mathcal{E}_{NM}^{f}$, are explicitely evidenced. 
   
The electron term is readily evaluated in the approximation of a degenerate, ultrarelativistic Fermi gas, %Because of the low electron mass, these approximations are valid also for very low
%densities. 
hence the electron chemical potential, $\mu_e$, is just equal to the electron Fermi momentum.  

The spin-saturated nuclear matter energy density, $\mathcal{E}_{NM}$, at finite temperature T, in the BCS approximation reads \cite{Chamel,Bur14}: 
%%
%\begin{eqnarray}
%\epsilon_{HM}(\rho_g,\delta_g)&=&g_q\sum_q \int_0^\infty \frac{dk}{2\pi^2} k^2 %f_q \frac{p^2}{2m^*_q}+\mathcal{E}_{sky}
%(\rho_g,\delta_g) \nonumber \\
%&+&\frac{1}{4}\sum_{q=n,p} v_\pi (\rho_{gq})\tilde{\rho}^*_{gq} \tilde{\rho}_{g%q} 
%\label{eq:HM}
%\end{eqnarray}
%
%\textcolor{red}{
%\begin{equation}
%\epsilon_{NM}(\rho,\delta) = g_i\sum_i \int_0^\infty \frac{dp}{2\pi^2 \hbar^3} p^2 f_i \frac{p^2}{2m^*_i}+\mathcal{E}_{pot},
%%(\rho_g,\delta_g) \nonumber \\
%%&+&\frac{1}{4}\sum_{q=n,p} v_\pi (\rho_{gq})\tilde{\rho}^*_{gq} \tilde{\rho}_{gq} 
%%\label{eq:HM}
%\end{equation}
%with
%\begin{equation}
%\mathcal{E}_{pot} = \mathcal{E}_{sky} + 
%\frac{1}{4}\sum_{i=n,p} v_\pi (\rho_{i})\tilde{\rho}^*_{i} \tilde{\rho}_{i} .
%\end{equation}
%%
%The pairing interaction acts only between nucleons of the same type $i$. }
\begin{equation}
%\epsilon_{NM} (\rho,y_p) = 2 \sum_i \int \frac{d\mathbf{p}}{(2\pi)^3} f_i \frac{p^2}{2m^*_i} + \mathcal{E}_{sky} + \frac{1}{4} \sum_{i=n,p} v_\pi (\rho_i) |\tilde{\rho}_i|^2.
\mathcal{E}_{NM} (\rho,y_p) = \sum_{i=n,p} \left [ 2 \int \frac{d\mathbf{p}}{(2\pi)^3} f_i \frac{p^2}{2m^*_i} +  \frac{1}{4}  v_\pi (\rho_i) |\tilde{\rho}_i|^2 \right ] + \mathcal{E}_{sky}.
\label{eq:HM}
\end{equation}
In the previous equation, $f_i$ is the occupation number for a nucleon of species $i$ with momentum ${\bf p}$ and $\tilde{\rho}_{i}=2\Delta_i(\rho_{i})/v_\pi(\rho_{i})$ denotes the anomalous 
density with the temperature dependent  pairing gap $\Delta_i$. The 
corresponding quasiparticle energies are given 
by $E_\Delta = \sqrt{\xi^2+\Delta_i^2}$, where $\xi = p^2/2m^*_i - \mu_i + U_i$, being $\mu_i$ and $U_i =\frac{\partial \mathcal{E}_{NM}}{\partial \rho_i}$ the chemical and mean field potential, respectively,  for each nucleonic species $i$.
%In the previous equation, $f_i$ is the 
%%Fermi function including pairing:
%particle occupation number given by:
%%
%\begin{equation}
%f_i=\frac {1}{2}  \left [ 1 - \frac{\xi}{E_\Delta} \tanh \left ( \frac{E_\Delta}{2 T}\right ) \r%ight ],
%\end{equation}
%%
%where $\xi = p^2/2m^*_i - \mu_i + U_i$ and $E_\Delta = \sqrt{\xi^2+\Delta_i^2}$, being $\mu_i$ a%nd $U_i =\frac{\partial \mathcal{E}_{NM}}{\partial \rho_i}$ the chemical and mean field potentia%l, respectively,  for each nucleonic species $i$.
%$\Delta_i$ is the temperature dependent  pairing gap and
%$\tilde{\rho}_{i}=2\Delta_i(\rho_{i})/v_\pi(\rho_{i})$ denotes the anomalous 
%density.  
We note that, in the definition of the mean-field potential $U_i$, the derivative with respect to  $\rho_{i}$
is taken at constant $\tilde{\rho}_{i}$.  
%$\Delta_i$ the temperature dependent pairing gap. It should be noticed that last term of the Eq.~\ref{eq:HM}, that is the pairing contribution to the potential part, depends also on the anomalous density $\tilde{\rho}_i = 2 \Delta_i (\rho_i) / v_\pi (\rho_i)$, which has to be taken constant making the derivative with respect to  $\rho_i$,  in the definition of the mean-field potential $U_i$.}
%It should be remarked that the pairing interaction  %(the last term in r.h.s. o%f Eq...)
%acts only between nucleons of the same type. 

In the following numerical applications,
we will use the SAMi-J35 parametrization \cite{Cha98} 
of the Skyrme energy functional for the local energy density 
$\mathcal{E}_{sky}$ 
and the effective nucleon mass $m^*_i$. 
For the pairing term, we adopt here the same interaction as in \cite{Chamel,Bur14}, whose density dependent strength,
$v_{\pi} (\rho_i)$, %by fixing the particle number density, 
is calculated exactly in the BCS approximation by inverting the gap equation, to reproduce 
the $^1S_0$ pairing gap of pure neutron matter given by Brueckner-Hartree-Fock calculations \cite{Cao06}.  
It should be remarked that, according to the asymmetry conditions of the stellar matter, the pairing interaction  %(the last term in r.h.s. of Eq...)
acts only between nucleons of the same type. 
The results are then extended to the $pp$ case, assuming that the pairing strength is the same as in the $nn$ case, just depending on the density of the species considered. 
Within this framework, one can then determine the derivatives of %\textcolor{red}{$\tilde{\mu}_i$ and $\Delta_i$} 
$\mu_i$ with respect to $\rho_i$ and so evaluate the curvature matrix 
\cite{Duc1},
which is needed in the calculation
of the cross section:
$$
\label{eq:matrix}
\mathcal{C} (q) =
\begin{pmatrix} 
\partial_{\rho_n} \mu_n & \partial_{\rho_ p} \mu_n & 0 \\
\partial_{\rho_n}\mu_p &  \partial_{\rho_p} \mu_p & 0 \\
      0 & 0 &  \partial_{\rho_e} \mu_e
\end{pmatrix}
+2 q^2 
\begin{pmatrix}
        C^f_{nn} & C^f_{np} & 0 \\
        C^f_{pn} & C^f_{pp} & 0 \\
        0 & 0 & 0
\end{pmatrix}
%\end{equation}
$$
\begin{equation}
+ \frac{4\pi e^2}{q^2}
\begin{pmatrix}
       0 & 0 & 0 \\
       0 & 1 & -1 \\
       0 & -1 & 1
\end{pmatrix} 
\label{eq:matrix}      
\end{equation}
where $e^2$ = 1.44~(MeV$\cdot$fm) and the coefficients  $C^f_{ij}$ are combinations
of the Skyrme surface parameters \cite{Duc1}.  
%\textcolor{blue}{Henceforth, we will make the choice to consider $\hbar = 1$.}
%E. Chabanat, et al., Nucl. Phys. A 627 (1997) 710â€“746.

%Since we deal with asymmetric matter, only p-p or n-n pairing will be considered.

\begin{figure}[t]
\includegraphics[scale=0.32]{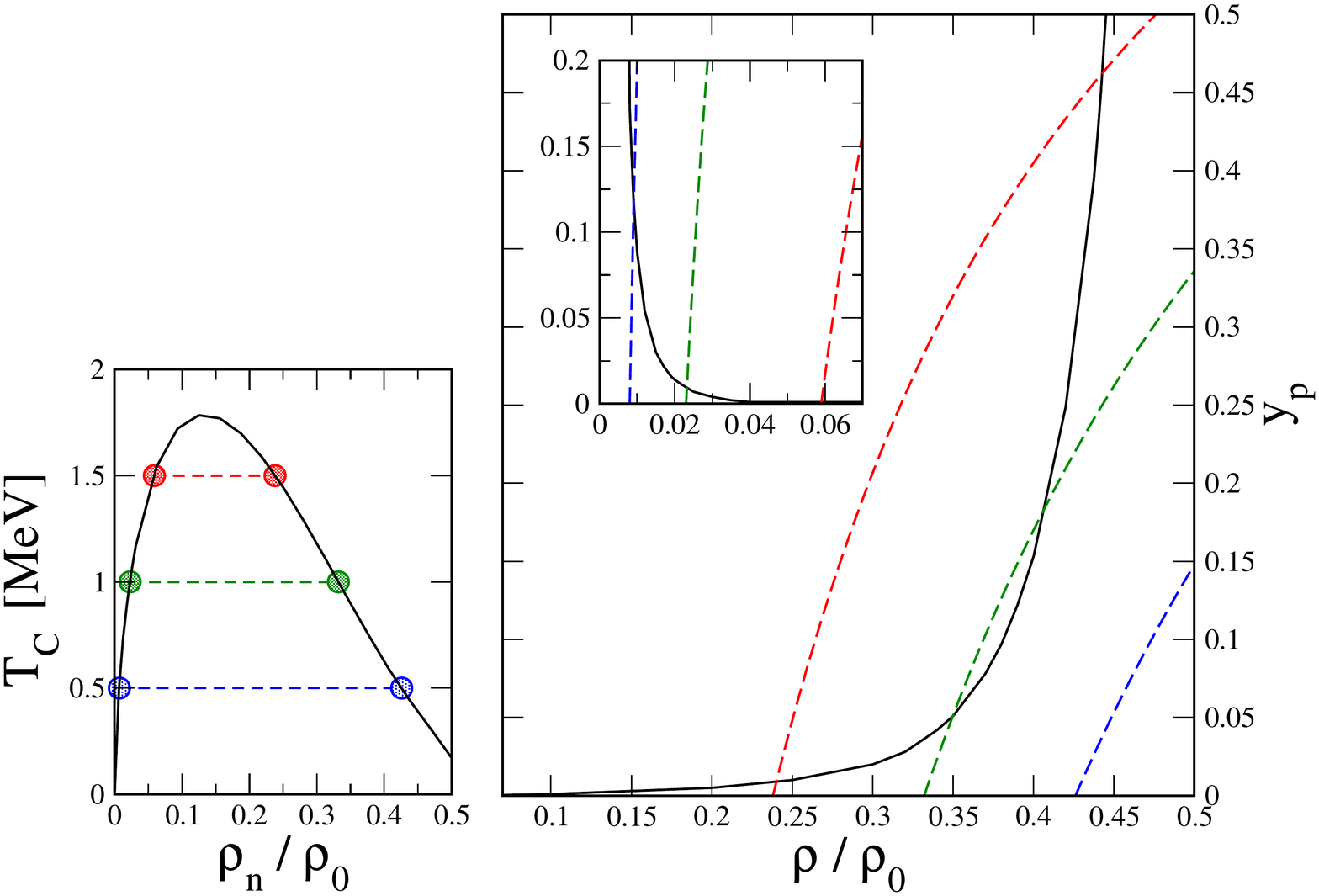} %,angle=270]
\caption{(Color online) Left panel: The critical temperature for the
transition from superfluid to normal matter, as a function
of the reduced neutron density $\rho_n/\rho_0$. 
Right panel: The spinodal border (full line), in the $(\rho,y_p)$ plane, 
associated with temperature T = 0.5 MeV and momentum transfer $q = 30$~MeV.
%$q = 0.15$~fm$^{-1}$. 
The inset shows a zoom of the low-density region. The dashed lines are curves of constant neutron density, 
%\textcolor{red}{corresponding to} \textcolor{blue}{wherein the critical temperature has} 
corresponding to the values associated with the circles
in the left panel (see text for more details).}  

\label{fig01} % optional figure label, must be unique
\end{figure}

\begin{figure}[t]
\includegraphics[scale=0.32]{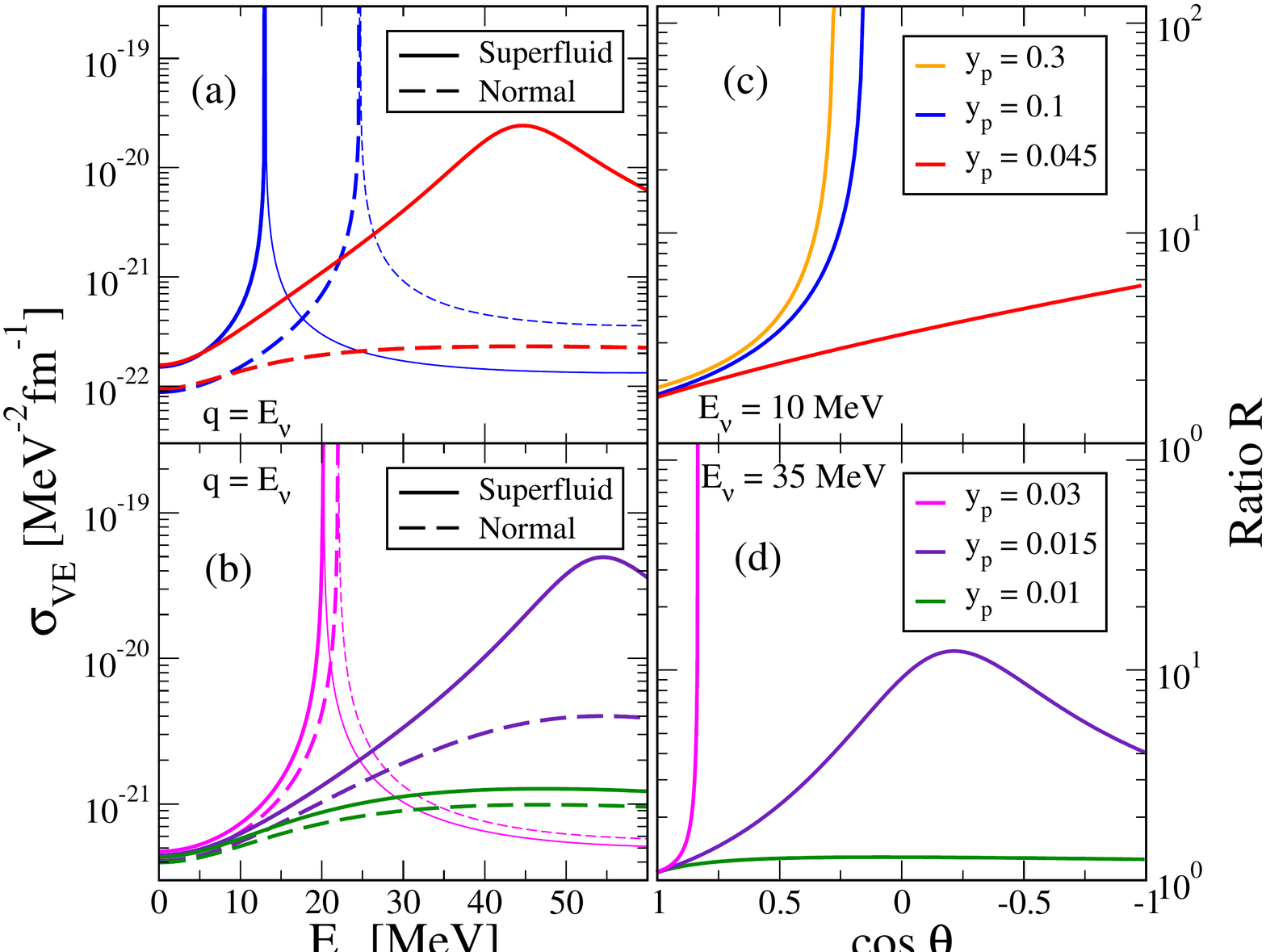}
\caption{(Color online)
Left panels: Neutrino differential cross sections, %per unit of volume 
$\sigma_{VE}$ (see text), as a function of the neutrino
energy, obtained in the full calculation (full lines) or neglecting the
pairing interaction (dashed lines). 
Right panels: Ratio $R$ between the full calculation and the results obtained
neglecting the pairing interaction, as a function of the cosine of the neutrino scattering
angle $\theta$, for selected neutrino energies. 
 Results are shown for the
following conditions: $\rho = \rho_0/100$ - T = 0.5 MeV (panels (a) and (c)) and 
 $\rho = \rho_0/4$ - T = 1.4 MeV (panels (b) and (d)). The proton fractions considered
are indicated inside the figure. 
}
\label{fig02} % optional figure label, must be unique
\end{figure}

\begin{figure}[t]
\includegraphics[scale=0.3]{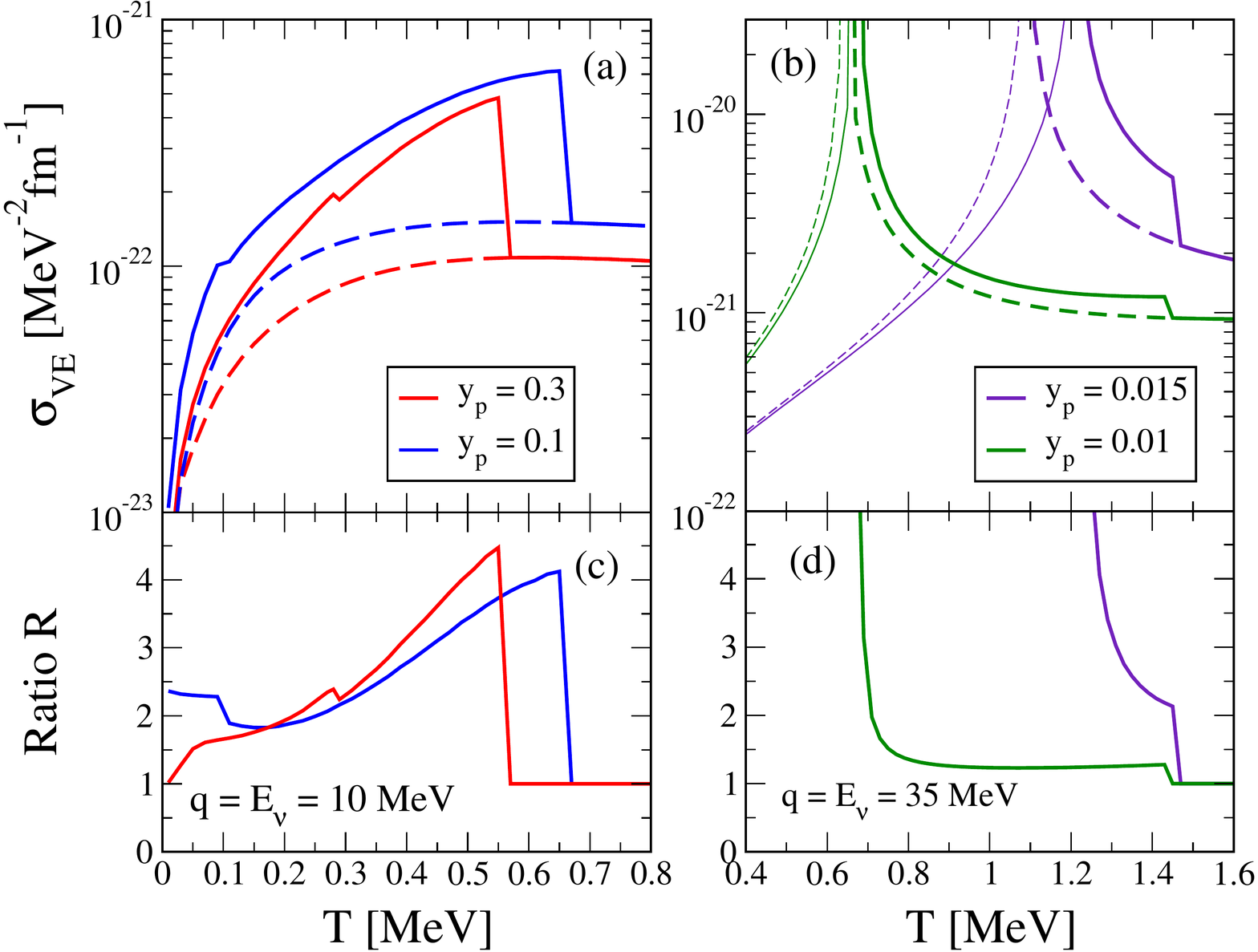}
%{cross_T_yp_infinity_new.ps}
\caption{(Color online) Top panels: 
Neutrino differential cross section, $\sigma_{VE}$ (see text), as a function of the temperature T, 
obtained in the full calculation (full lines) or neglecting the pairing interaction (dashed lines). 
Bottom panels: Ratio $R$ between the full calculation and the results obtained
neglecting the pairing interaction, as a function of the temperature T. Results are shown for the 
following conditions:  $\rho = \rho_0/100$ (panels (a) and (c)) and $\rho = \rho_0/4$ (panels (b) and (d)). 
The proton fractions considered are indicated inside the figure. }
\label{fig03} % optional figure label, must be unique
\end{figure}

%The goal of our analysis is to assess the role of pairing effects 
%on neutrino transport in neutron-rich nuclear matter.  
The pairing interaction modifies neutron and
proton chemical potentials, and their derivatives, which appear in the
curvature matrix, Eq.~\eqref{eq:matrix}. Thus, in suitable conditions of density, asymmetry
and temperature, we can expect a non--negligible impact on the neutrino
differential cross section.   

As stressed before, neutrino trapping is quite influenced
by large density fluctuations of the nuclear density, which develop
close to the spinodal border and may lead to clustering phenomena. 
%as those occurring in supernovae explosion and neutron stars~\cite{Bur15}}.
Within the framework 
adopted here (see Eq.(5)), the amplitude of
neutron and proton density fluctuations is essentially related to the
inverse of the eigenvalues of the curvature matrix  and 
%\textcolor{red}{, in particular, to the
%isoscalar-like, $\lambda_s(q)$, and isovector-like, $\lambda_v(q)$ eigenvalues, becoming}
so become quite large when the isoscalar-like one, $\lambda_S (q)$, is small. 
%This happens in particular close to the spinodal border,  where $\lambda_s$ van%ishes. %\textcolor{red}{and then, inside the instability region, becomes negati%ve.} 
%%\textcolor{blue}{
In this case, %when $\lambda_s$ is small, 
we expect pairing correlations
to have a large relative weight on the curvature matrix elements, especially close
to the critical temperature, $T_c$,  for the transition from normal to superfluid matter, where discontinuities appear in the chemical potential derivatives \cite{Bur14}.
%Since the spinodal region is associated with mechanical instabilities occurring
%at low density and leading to matter clusterization, 
%%\textcolor{red}{. Then} 
%it is certainly
%a relevant density region %\textcolor{red}{of relevance} 
%when discussing neutron stars and supernova
%explosions, where low-density clustering occurs~\cite{Bur14}. 
%\textcolor{red}{However, when $\lambda_s$ is small, we expect pairing correlations
%to have a large relative weight on the curvature matrix elements, especially close
%to the critical temperature, $T_c$,  for the transition from normal to superfluid m%atter.}
Hence we will discuss stellar matter at moderate temperature (below 2 MeV), at
density and asymmetry conditions close to the spinodal border, where $\lambda_S$ vanishes.  
%Moreover, for a given neutron or proton density values, pairing effects
%are quite large close to the critical temperature, $T_c$,  for the transition from normal to superfluid matter.   
%Here we consider neutron-rich matter and pairing effects are mostly important
%in the n-n channel. 
%%%%%  le fluttuazioni sono essenzialmente determinate dall'autovalore
%%%%%%  piu' piccolo, cioe' \lambda_s

For the sake of simplicity, let us start considering only neutron pairing. 
As shown in the following, the latter leads, in any case, to the largest effects.    
%Therefore, in order to better isolate the temperature, density and asymmetry conditions
%where pairing effects may give an important contribution, 
In Fig. 1 (left panel) we represent the critical temperature, $T_c$  as a function of the
neutron density.  
%So we will concentrate our analysis on temperature, density and asymmetry
%conditions which are close to the spinodal border and also close to the
%pairing critical temperature (for neutrons). 
%In Fig.1 we represent the critical temperature as a function of the neutron
%density.  
The right panel shows the spinodal border (full line), in the ($\rho, y_p$) plane, at a %typical 
temperature and $q$ values of interest for our study. It should be noticed that, because of 
Coulomb and surface effects, the spinodal border depends on $q$,
%\textcolor{red}{. On the other hand,} \textcolor{blue}{, although} 
but  it is not very sensitive to   %it does not evolve much with 
the temperature, within the range considered in our study.
For a fixed value of the temperature T, the plot on the left panel allows one
to identify two values of the neutron density $\rho_n$ (see the circles) 
for which the temperature considered corresponds to the critical one, so pairing effects are
expected to be large. Each value of the neutron density
%\textcolor{red}{, $\rho_n = \rho (1-y_p)$,} 
defines 
%\textcolor{red}{a curve} \textcolor{blue}
{an hyperbole} in the ($\rho, y_p$) plane (see the dashed lines on the right panel).
Then the crossing of the dashed lines with the spinodal border identifies
the density-asymmetry regions where 
large density fluctuations can coexist with important pairing contributions.
It appears that a variety of conditions, from very small densities up to
$\rho \approx 0.4 \rho_0$ (being  $\rho_0 = 0.16$ fm$^{-3}$ the saturation density) and with proton fraction ranging from quite low
values up to $y_p\approx 0.5$, 
are good candidates for our study.   These conditions may occur in the inner crust
of a neutron star or in the pre-bounce phase of a supernova explosion,
when the temperature is still low (see e.g. refs.\cite{Ma87,Pra01,Pa12,Buyuk,Bur15}).    
%\textcolor{green}{Maybe should we cite here the paper on the inner crust with Gulminelli?}
    
%At higher temperature (see the result at T = 1.5 MeV for instance) also
%regions at larger density and proton fraction can contribute to our analysis
%(see the crossing with the red curve on the right panel).  This region
%can be of interest for the description of supernova explosion. 

Guided by this analysis, 
in the following we will discuss the results obtained 
for the two sets of parameters: (1) $\rho = \rho_0/100$, at T= 0.5 MeV, 
and (2) $\rho = \rho_0/4$, at T= 1.4 MeV, %with $\rho_0 = 0.16$ fm$^{-3}$, 
i.e. we will consider two opposite density regimes at temperatures where we expect 
large pairing effects. 
Several $y_p$ values will be tested, close to the values suggested by Fig.1.
For the conditions considered now, only neutrons are paired, because the temperature values are always above
the proton critical temperature.
%{\bf The first set of parameters may correspond to a particular step of the pre--bounce phase of a 
%supernova, while the latter set may represent a physical situation for the inner crust of a 
%neutron star.} 

In Fig.2 (left panel) we display the neutrino differential cross section,
$\sigma_{V E} \equiv 1/(VE_\nu^2)~ d^2\sigma/d\Omega^2$ (full lines), as a function of the neutrino energy $E_\nu$ (we
consider momentum transfer
$q = E_\nu$).  To underline the importance of pairing effects, we also show 
(dashed lines) the results obtained neglecting the
pairing interaction in our calculations. 
%whereas full lines correspond to the complete calculations.  
At the lowest density considered (case (1), panel (a)), the proton fraction $y_p = 0.045$ corresponds
to conditions which are close, but outside the spinodal region for all $q$ values. One can notice 
quite large pairing effects on the cross section, especially for intermediate neutrino energies, where a bump is 
observed in the full calculation. In this case the momentum transfer corresponds to 
density oscillations %associated with $q$ values
not much affected by Coulomb (acting at small $q$s) or surface (acting at
large $q$s) effects, so that $\lambda_S$ remains close to zero for an appreciable extent 
of the momentum transfer and the pairing effects are clearly enhanced. 
%for which $\lambda_s$ becomes close to zero and pairing effects are clearly important.   
We notice that pairing correlations go in the direction of reducing the curvature of the free energy density,
leading to an increase of density fluctuations and related neutrino cross section.
This effect is associated with neutron pairing and indicates that neutron correlations favours matter
clustering. 

Due to Coulomb repulsion effects, when the momentum transfer $q$ approaches zero, the eigenvalue $\lambda_S (q)$ is always positive and therefore density oscillations are stable. 
However, for higher proton fractions (see the result for $y_p = 0.1$), 
nuclear matter approaches the spinodal border already for oscillations
related to quite low momentum transfer. %$q$ values. 
At the crossing, a divergent behavior is observed for density fluctuations and neutrino
cross section. This is reminiscent of the well--known critical--opalescence phenomenon 
observed in the light 
scattering through a fluid near the critical temperature of the liquid--vapor transition \cite{La84}. 
Clearly here one should go beyond the second curvature
of the free energy density, in order to evaluate density fluctuations accurately. 
However, our calculations already point out strong pairing effects, with a significant shift to smaller values of the neutrino energy associated with the divergency in the full calculations.
This implies that also less energetic neutrinos have more chances to be trapped, so that the energy 
flux carried away by neutrinos could be damped by pairing correlations.

Larger transfer momenta $q$s correspond to unstable oscillations.  In this case $\lambda_S (q)$ is
negative and one cannot simply apply the prescription given above (Eq.(7))
to evaluate neutron and proton density fluctuations.
Indeed, whereas in stable situations the variance associated with isoscalar-like fluctuations equals
$\sigma_S(q) = T/\lambda_S(q)$, in presence of instabilities it %fluctuations
grows exponentially with time until a new equilibrium condition, corresponding to 
clustered matter, is reached \cite{rep04}. 
%according to the expression (obtained 
%Within the linear response regime, the initial fluctuation growth can be expres%sed as:
%\begin{equation}
%\sigma_s(q)= \frac{T}{|\lambda_s(q)|} (e^{2t/\tau(q)} - 1),
%\end{equation}        
%where $\tau(q)$ denotes the instability growth time of the $q$ mode considered. 
%In stellar matter we are dealing with processes associated with quite long time% scales, 
%thus surely beyond the linear response regime. 
Hence, inside the spinodal region the
correct equilibrium fluctuations cannot be estimated  within our framework.
However, as a  first order approximation, we assume that the equilibrium
variance is close to the value obtained, for each $q$, at the time
$t\approx \tau(q)$, being  $\tau(q)$ the instability growth time of the $q$ mode considered \cite{Comat},  
%(as observed in nuclear multifragmentation)
leading to  $\sigma_S(q) \approx {T}/{|\lambda_S|}$. %\frac{T}{|\lambda_S|}$
Then the elements of the curvature matrix, Eq.(7), are modified accordingly. 
%For the sake of simplicity, and in order to preserve an analogy with the
%stable case, we just take $\sigma^s_q = \frac{T}{|\lambda_s|}$, neglecting
%the exponential increase, as a representative value of fluctuations inside the
%spinodal region, keeping in mind that this does not represent the
%equilibrium varinace in this case.  
The corresponding neutrino cross sections 
%obtained, in this way, for unstable $q$ values 
are indicated by thin lines in Fig.2.  

The panel (b) displays the results obtained for case (2), 
%at $\rho = \rho_0/4$ - T= 1.4 MeV.
where, as indicated by the analysis shown in Fig.1, we take smaller proton fractions $y_p$. We note that this also corresponds to the trend predicted for
the proton fraction in the inner crust of neutron stars \cite{Camille,Bur15}.  
The same considerations made above for the lower density case hold. 
However here pairing effects, though still quite significant, are reduced with respect to 
the previous case, just because they are linked to the derivative of the 
pairing gap (and thus of the critical temperature) with respect to the density \cite{Bur14},
which is steeper for case (1) (see Fig.~\ref{fig01}).  

To emphasize the role of pairing effects,
panels (c),(d) of Fig.\ref{fig02} represent the ratio $R$ between the cross section associated with the full calculations
and the results obtained neglecting the pairing interaction, as a 
function of $\cos \theta$,
for selected neutrino energies,  
%The chosen value of neutrino energy is 
representative of $\beta$--equilibrium conditions.
We note that the neutrino scattering angle is related to 
the momentum transfer $q$.
% which is closely related to the momentum transfer $q$. 
Unstable $q$ values are not considered in this plot. 
Results are shown for three proton fractions, for cases (1) and (2).  
From this representation, it clearly emerges how huge pairing effects become 
approaching the spinodal border.  
%%%%  pairing is negative, it reduces the \lambda.

The influence of the temperature on our results is discussed in Fig.~\ref{fig03},
where the quantity $\sigma_{VE}$
%$\sigma_{VET} \equiv 1/T~\sigma_{VE}$ 
is displayed, for selected neutrino energies and $q = E_\nu$, as a function of T, for 
density conditions as in case (1) (panel (a)) and case (2) (panel (b)). Two proton fractions
are considered.  
The full calculations are compared to those obtained neglecting the 
pairing interaction. 
The momentum transfer considered in panel (a) corresponds to stable oscillations.
Pairing effects are quite important already at very low temperature, where
$pp$ pairing is also present, but they increase
approaching the neutron critical temperature, $T_c\approx 0.65 (0.55)$ MeV for
$y_p = 0.1 (0.3)$ , and then vanish. 
Indeed, quite interestingly, we observe a jump in the cross section at $T = T_c$, which suddenly
reaches the value of normal nuclear matter.  A small jump also occurs, for 
both proton fractions, at a
lower temperature, %$T \approx 0.05-0.1$ MeV, 
due to the disappearance of proton pairing. 
The jumps observed are related to the discontinuity 
%also manifest 
emerging in the density derivative of the chemical
potential, $\partial \mu_i / \partial \rho_i$, which is connected to the matter compressibility,
in analogy with the well known heat capacity discontinuity \cite{Bur14,Bur15}. 
%\textcolor{green}{Maybe should we cite here the paper on the inner crust with Gulminelli?}
%!!!   occurring at the critical temperature. 
     
The conditions of panel (b) of Fig.\ref{fig03} are such that the $q$ value considered corresponds
to fluctuations which are unstable at zero temperature.  
%When increasing the temperature, 
A divergency occurs for the cross sections at the temperature associated with the crossing of the spinodal border. Then, at higher temperature, 
density oscillations become stable. We note that, 
inspite of the increasing temperature, the cross section reduces when
the nuclear matter moves away from the spinodal border.  
Also in this case %the largest pairing effects emerge at the neutron critical temperature, where
a discontinuity is observed at the neutron critical temperature.
To better evidence the role of pairing effects, 
panels (c) and (d) of Fig.3 
show the ratio $R$ between the full calculations and the results obtained neglecting pairing correlations. 
%In both cases A small jump (not shown) occurs at the proton critical temperature, as for the calculations displayed in the left panel. 

To conclude, our analysis evidences important pairing effects on neutrino transport, for suitable density, asymmetry
and temperature conditions, which are of relevant interest for the evolution of neutron stars and supernovae explosion in the pre-bounce phase \cite{Buyuk,Bur15}.
We concentrate on the behavior of low-density matter, close to the spinodal border, characterized by quite
large density fluctuations.   
Focusing on neutral current neutrino scattering, we generally observe
an increase of the neutrino differential cross section in paired matter, thus enhancing neutrino trapping and reducing the energy flux carried out by neutrino
emission. 
This is essentially due to attractive neutron-neutron pairing correlations, which favour low-density clustering. 
From this study new hints emerge about 
%Our results indicate that pairing effects can have 
a significant impact of pairing effects on the cooling mechanism, by neutrino
emission, of low-density stellar matter at moderate temperature.


\begin{thebibliography}{}
%Nucleon effective masses within the Brueckner-Hartree-Fock theory: Impact on stellar neutrino emission
%By: Baldo, M.; Burgio, G. F.; Schulze, H. -J.; et al.
%PHYSICAL REVIEW C   Volume: 89   Issue: 4     Article Number: 048801   Published: APR 28 2014 

%Sanjay Reddy, Madappa Prakash, James M. Lattimer, and Jose A. Pons    PHYSICAL REVIEW C MAY 1999 VOLUME 59, 2888 




\bibitem{Be85} H. A. Bethe and J. R. Wilson, Astrophys. J. {\bf 295}, 14 (1985). 
\bibitem{Bu90} A. Burrows, Annu. Rev. Nucl. Part. Sci. {\bf 40}, 181 (1990).
\bibitem{Pra01} M. Prakash, J. M. Lattimer, J. A. Pons, A. W. Steiner, and S. Reddy, in
{\it Lectures Notes in Physics}, edited by D. Blaschke, N. K.  Glendenning, and 
A. Sedrakian (Springer--Verlag, Berlin, 2001) Vol. 578, p. 364. 
\bibitem{Ja07} H. T. Janka, K. Langanke, A. Marek, G. Mart\'inez--Pinedo, and B. Mueller, 
Phys. Rep. {\bf 442}, 38 (2007). 
\bibitem{Bal14} M. Baldo {\it et al.}, %M.; Burgio, G. F.; Schulze, H. -J.; et al.
Phys. Rev. C {\bf 89}, 048801 (2014).
\bibitem{Re98} S. Reddy, M. Prakash, and J. M. Lattimer, Phys. Rev. D {\bf 58}, 013009 (1998). 
\bibitem{Re99} S. Reddy, M. Prakash, J. M. Lattimer, and J. A. Pons,    
Phys. Rev. C {\bf 59}, 2888 (1999). 
\bibitem{Bu98} A. Burrows and R. F. Sawyer, Phys. Rev. C {\bf 58}, 554 (1998). 
\bibitem{Re00} S. Reddy, G. F. Bertsch, and M. Prakash, Phys. Lett. B {\bf 475}, 1 (2000). 
\bibitem{Wa03} G. Watanabe, K. Sato, K. Yasuoka, and T. Ebisuzaki. Phys. Rev. C {\bf 68}, 
035806 (2003). 
\bibitem{Ho04} C. J. Horowitz, M. A. P\'erez--Garc\'ia, and J. Piekarewicz, Phys. Rev. C {\bf 69}, 
045804 (2004). 
\bibitem{Ma12} G. Mart\'inez--Pinedo, T. Fischer, A. Lohs, and L. Huther, Phys. Rev. Lett. 
{\bf 109}, 251104 (2012). 
\bibitem{Pa14} H. Pais, W. G. Newton, and J. R. Stone, Phys. Rev. C {\bf 90}, 065802 (2014). 
\bibitem{rep04} Ph. Chomaz, M. Colonna,  and J. Randrup, Phys. Rep. {\bf 389}, 263 (2004).
\bibitem{Borderie}B. Borderie and M. F. Rivet,
Progr. in Part. and Nucl. Phys. {\bf 61} (Book Series), 551 (2008), and refs. therein.  
\bibitem{Lattimer}J. M. Lattimer and M. Prakash, Phys. Rep. \textbf{442}, 109 (2007);
%Title: THE EQUATION OF STATE FROM OBSERVED MASSES AND RADII OF NEUTRON STARS
A. W. Steiner, J. M.Lattimer, and E. F. Brown, 
Astrophysical J.  {\bf 722}, 33 (2010).  
\bibitem{Camille}Ad. R. Raduta, F. Gulminelli, and F. Aymard, Eur. Phys. J. {\bf A50}, 24 (2014).
%F. Gulminelli, Physica Scripta {\bf T152}, 014009 (2013).
\bibitem{Marg2004} J. Margueron, J. Navarro, and P. Blottiau, Phys. Rev. C 70, 028801 (2004).
\bibitem{Pa09} D.Page, J. M. Lattimer, M. Prakash, and A. W. Steiner, Astrophys. J. {\bf 707}, 
1131 (2009). 
\bibitem{Ca00} B. Carter, in {\it Lectures Notes in Physics}, edited by D. Blaschke, N. K. Glendenning, 
and A. Sedrakian (Springer--Verlag, Berlin, 2001) Vol. 578, p. 54. 
\bibitem{Se05} A. Sedrakian, Phys. Rev. D {\bf 71}, 083003 (2005). 
\bibitem{Bur14} S. Burrello, M. Colonna, and F. Matera,  Phys. Rev. C {\bf 89}, 057604 (2014).
\bibitem{Iw82} N. Iwamoto and C. J. Pethick, Phys. Rev. D {\bf 25}, 313 (1982).
\bibitem{Landau}L. D. Landau and E. M. Lifshitz, {\it Statistical Physics Part 1}, 
Vol. 5 (3 ed.) (Butterworth-Heinemann, Oxford, 1980). 
%ISBN 978-0-750-63372-7 (1980)
%\bibitem{Ciccio_ref}
\bibitem{Chamel}N. Chamel, Phys. Rev. C {\bf 82}, 014313 (2010). 
\bibitem{Cha98} X. Roca-Maza {\it et al.}, Phys. Rev. C {\bf 87}, 034301 (2013). %X. Roca-Maza, G. Col\`{o}, H. Sagawa, 
%Phys. Rev. C {\bf 86}, 031306(R) (2012). 
%E. Chabanat, P. Bonche, P. Haensel, J. Meyer, and R. Schaeffer, Nucl.  Phys.  
%A {\bf 635}, 236 (1998). 
\bibitem{Cao06} L.  G. Cao, U. Lombardo, and P. Schuck , Phys. Rev. C {\bf 74}, 064301 (2006). 
\bibitem{Duc1}C. Ducoin, Ph. Chomaz, and F. Gulminelli, Nucl. Phys. A {\bf 789}, 403 (2007).
\bibitem{Ma87} R. Mayle, J. R. Wilson, and D. N. Schramm, Astrophys. J. {\bf 318}, 288 (1987).
\bibitem{Pa12} D. Page and S. Reddy, in {\it Neutron Star Crust}, edited by C. A. Bertulani and 
J. Piekarewicz (Nova Science Publishers, New York, 2012), p. 281. 
\bibitem{Buyuk} N. Buyukcizmeci et al., Nucl. Phys. A {\bf 907}, 13 (2013). 
\bibitem{Bur15}S. Burrello {\it et al.}, Phys. Rev. C {\bf 92}, 055804 (2015).
\bibitem{La84} L. D. Landau, E. M. Lifshitz, and L. P. Pitaevskii, {\it Electrodynamics of 
Continuous Media}, Vol. 8 (2 ed.) (Butterworth--Heinemann, Oxford, 1984). 
\bibitem{Comat} M. Colonna and F. Matera, Phys. Rev. C {\bf 71}, 064605 (2005);
 Phys. Rev. C {\bf 77}, 064606 (2008). 



\end{thebibliography}
\end{document}